\documentclass[nobibnotes,aps,superscriptaddress,10pt,notitlepage,twocolumn,longbibliography,nofootinbib]{revtex4-1}

\usepackage{comment}
\usepackage[latin1]{inputenc}
\usepackage{graphicx}
\graphicspath{{Figures/}}
\usepackage{float}
\usepackage{latexsym}
\usepackage{graphicx}
\usepackage{amssymb}
\usepackage{amsmath}
\usepackage{amsfonts}
\usepackage[colorlinks=true,citecolor=blue,hyperfootnotes=false]{hyperref}
\usepackage[dvipsnames]{xcolor}
\usepackage{cool}
\usepackage{dcolumn}
\usepackage{textcomp}
\usepackage{xcolor}
\usepackage{xfrac}
\usepackage{slashed}
\usepackage{multirow}
\usepackage[scr=esstix]{mathalpha}

\def\be{\begin{equation}}
\def\ee{\end{equation}}
\def\beq{\begin{equation}}
\def\eeq{\end{equation}}
\def\figs/B{B}
\def\bea{\begin{eqnarray}}
\def\eea{\end{eqnarray}}
\def\bg{\begin{eqnarray}}
\def\nd{\end{eqnarray}}

\newcommand{\bx}{\boldsymbol{x}}
\newcommand{\bk}{\boldsymbol{k}}

\def\cos{{\rm cos}}

\newcommand{\ma}{m_A}
\newcommand{\na}{n_A}
\newcommand{\rhoa}{\rho_A}
\newcommand{\mh}{m_h}
\newcommand{\nh}{n_h}
\newcommand{\rhoh}{\rho_h}
\newcommand{\kJh}{k_{G,h }}
\newcommand{\kJa}{k_{G,A }}
\newcommand{\csh}{c_{s,h}}
\newcommand{\h}{\mathscr{h}}
\newcommand{\A}{\mathcal{A}}

\def\be{\begin{equation}}
\def\ee{\end{equation}}

\def\cos{{\rm cos}}

\begin{document}

\title{Cold and Fuzzy Dark Sector}

\author{Christian Capanelli}
\affiliation{Department of Physics \& Trottier Space Institute,
McGill University, Montr\'eal, QC H3A 2T8, Canada}

\author{Elisa G. M. Ferreira}
\affiliation{Kavli Institute for the Physics and Mathematics of the Universe (WPI),
UTIAS, The University of Tokyo, Chiba 277-8583, Japan}

\author{Evan McDonough}
\affiliation{Department of Physics, University of Winnipeg, Winnipeg, MB R3B 2E9, Canada}

\begin{abstract}
  We introduce the Fuzzy Dark Sector (FDS) scenario as a rich, interacting system and candidate for dark matter. This serves as a natural extension of the single-component, non-interacting Fuzzy Dark Matter (FDM) paradigm.  Concretely, we consider an ultra-light Abelian-Higgs model, with interacting Higgs and dark photon degrees of freedom. We find that the transfer function, and hence imprint on the CMB and Large-Scale Structure (LSS), is characterized by a single characteristic scale of the interacting fuzzy dark sector, allowing us to recover the LSS signature of single-field FDM, dependent on the FDS parameters. In contrast, galactic halos present a great {\it diversity}, unlike with the universality of single-field FDM, owing to the interaction between fields. This interaction introduces an instability that is not otherwise present for the case of four decoupled scalars. Finally, we comment on primordial production and portals to the Standard Model, and introduce another simple realization of the Fuzzy Dark Sector paradigm with a kinetic coupling. 
\end{abstract}

\maketitle

\tableofcontents

\section{Introduction}\label{sec:intro}

The conventional, single component, non-interacting fuzzy dark matter (FDM) \cite{Hu:2000ke} scenario provides a novel and elegant explanation for dark matter.

This model has been heavily tested in the past few years, using diverse probes from large and small-scale observations, with the allowed mass range for this candidate being systematically narrowed down \cite{Ferreira_2021,Hui:2021tkt, Eberhardt:2025caq,Rogers:2020ltq, Dalal:2022rmp,Powell:2023jns,Liu:2025cxq}. Both from observational constraints and theoretical considerations, it is natural to extend the FDM paradigm to multiple interacting species, hence exploring the \textit{zoo} of ultra-light dark matter models.

Multiple ultra-light degrees of freedom emerge naturally from a variety of UV completions. Perhaps the best known example is the string axiverse \cite{Svrcek:2006yi,Arvanitaki:2009fg, Cicoli:2012sz, Demirtas:2021gsq,Sheridan:2024vtt}, where a spectrum of light axions emerge with roughly log-distributed masses, although alternatives also exist in even the context of field theory \cite{Maleknejad:2022gyf,Alexander:2023wgk,Jain:2023ojg,Alexander:2024nvi,Alexander:2025flo}. Aside from axions, a prominent example of interacting ultra-light dark sectors in string theory is the axio-dilaton, see e.g. \cite{Burgess:2021juk,Burgess:2021obw}. The phenomenology of multiple ultra-light scalars has been considered in cosmology \cite{Lague:2021frh,Lague:2023wes, Tellez-Tovar:2021mge}, typically assuming they comprise only some sub-fraction compared to CDM. In this setting, the addition of extra degrees of freedom allows for greater diversity in halos \cite{Luu:2024lfq}. Similarly, the small-scale structure of non-interacting ultra-light vector and higher spin DM has been studied, which can be mapped to a set of non-interacting scalars \cite{Jain:2021pnk,Gorghetto:2022sue,Amin:2022pzv,Nomura:2025jur}.

On the other hand, self-interacting fuzzy dark matter (SIFDM) has re-gained recent interest as a natural extension of the standard FDM model\cite{2011PhRvD..84d3531C,Chavanis:2017loo,Suarez:2017mav,Chakrabarti:2022owq,Dave:2023egr, Dave:2023wjq, Mocz:2023adf} and for its unique phenomenology \cite{Glennon:2022huu,Glennon:2023gfm,Capanelli:2025nrj, Indjin:2025thr}. Most recently, it was shown that a rich diversity of solitons can be achieved for FDM with multiple interacting components \cite{Glennon:2023jsp,Mirasola:2024pmw,Toomey:2025mvx}. Given the successes of these extensions, both adding a self-interaction and coupling multiple ultra-light fields, it is natural then to look for concrete model realizations.

In this work, we propose a \textit{fuzzy dark sector} (FDS) comprised of an ultra-light Abelian-Higgs model. Our model features both a Higgs field with repulsive self-interaction and a massive dark photon---giving an additional three degrees of freedom with a shared mass and that couple directly to the Higgs.

 In this work we explore the cosmology of the FDS. We find that unlike decoupled multicomponent FDM, the FDS has a \textit{single} characteristic collapse scale. As a consequence, we find the matter power spectrum in the FDS scenario mimics that of non-interacting single-component FDM, with a parametric freedom to vary the apparent FDM mass: the FDS can mimic a much heavier or much lighter FDM candidate, depending on the FDS parameters.

A study of the solitons of the FDS, which would constitute cores in dark matter halos, reveals a striking diversity, with density profiles depending on the relative mass fraction in the different components of the FDS and the interaction strengths. Solitons provide a direct probe of the interactions of the FDS: 
We find that the attractive interaction between the Higgs and gauge field leads to a critical soliton mass beyond which solitons are unstable, and which would not be present in the non-interacting theory.

The outline of this paper is as follows: In Sec.~\ref{sec:model} we introduce the model, deriving the non-relativistic equations of motion, and then formulate the effective description in terms of fluid variables. In Sec.~\ref{sec:jeans}, we study FDS cosmology, performing a Jeans-type analysis and computing the primordial matter power spectrum. In Sec.~\ref{sec:solitons}, we use variational methods to estimate the size and stability of solitons, finding a diversity of halo profiles. Finally, in Sec.~\ref{sec:discussion}, we discuss the possible production mechanisms for the FDS, comment on possible portals between the FDS and Standard Model, and introduce another concrete realization of a Fuzzy Dark Sector.

\section{Model}\label{sec:model}

In this work, we focus on the Abelian-Higgs model, defined by the relativistic action,
\begin{eqnarray}
\label{eq:AbelianHiggs}
S=\int \mathrm{d}^4 x \sqrt{-g} && \left[\right.\frac{R}{16 \pi G}-\frac{1}{4} F_{\mu \nu} F^{\mu \nu}+\frac{1}{2}  |D_{\mu}\phi|^2 \nonumber \\
&&-\frac{\lambda}{4!}\left(|\phi|^2-v^2\right)^2\left.\right]\,. 
\end{eqnarray}
This describes a complex scalar $\phi$ (the ``Higgs") and an Abelian gauge field $A_\mu$, interacting through the covariant derivative $D_{\mu}=\partial_{\mu}+i g A_{\mu}$ with gauge coupling $g$. 

At low energies, the scalar exhibits spontaneous symmetry breaking, where the gauge symmetry is non-linearly realised, and $\langle \phi \rangle =v$. In this phase of the theory, the field can be expanded as $\phi =(h+v)e^{i\theta}$ and the action takes the form (in unitary gauge $\theta=0$)
\begin{eqnarray}
    S=&&\int \mathrm{d}^4 x \sqrt{-g}\left[\right.\frac{R}{16 \pi G}-\frac{1}{4} F_{\alpha \beta} F^{\alpha \beta}+\frac{1}{2}  (\partial_{\mu }h)^2\\
    &&
    +\frac{\ma^2}{2}A_{\mu}^2-\frac{\mh^2}{2} h^2+g^2v A_{\mu}^2 h + \frac{g^2}{2}A_{\mu}^2h^2 -\frac{\lambda}{4!}h^4 \left.\right] \nonumber \,,
\end{eqnarray}
where the masses are given by
\begin{eqnarray}
    \ma^2 &&\equiv g^2 v^2 , \\ 
    \mh^2&&\equiv 2\lambda v^2 .
\end{eqnarray}
We are interested in the possibility that fuzzy dark matter is realized in this phase of the Abelian Higgs model, where we identify $h$ and $A_{\mu}$ as the scalar and vector fuzzy dark matter components of a fuzzy dark sector. 

By analogy to the axion and axion-like particles, which are described by the same action Eq.~\eqref{eq:AbelianHiggs} but with $v\equiv f_a$ and without the gauge field, we identify the Higgs VEV $v$ as a UV scale in the model, namely
\begin{equation}
    v \,\, ({\rm Abelian-Higgs}) \leftrightarrow f_{a} \,\, ({\rm Axion}).
\end{equation}
 We realize an ultra-light scale of fuzzy\footnote{Although the label ``fuzzy" is often reserved for particle masses $m\sim10^{-22} \ {\rm eV}$, we will henceforth refer to any generalization where the de Broglie wavelength is astrophysically relevant as fuzzy.} or wave dark matter, $m\ll {\rm eV}$, through extremely weak interactions $\lambda$ and $g$,
\begin{equation}
    m_{a},m_h \ll {\rm eV} \Rightarrow \lambda, g \ll 1\,.
\end{equation} 
 Thus, the Fuzzy Dark Sector is necessarily comprised of feebly interacting massive particles (FIMPs), as reviewed in e.g.~\cite{Hall:2009bx,Chu:2011be,Bernal:2017kxu,Lanfranchi:2020crw}. Motivations for such small couplings range from technical naturalness to approximate accidental symmetries. Here we remain agnostic to the mechanism to explain the smallness of the couplings.

The wave regime of the Fuzzy Dark Sector emerges in the non-relativistic limit of the theory. The non-relativistic limit for ultra-light vector dark matter has been studied in detail in \cite{Jain:2021pnk,Amin:2022pzv}. Here we generalize that analysis to the Abelian-Higgs model.

\subsection{Relativistic Action for Physical Degrees of Freedom}

We first rewrite the relativistic theory in terms of physical degrees of freedom. We decompose $A_{\mu}$ into temporal and spatial components $A_0$ and $A_i$. The former satisfies the constraint equation
\begin{equation}
\left(\nabla^2 + m^2 _{A,{\rm eff}}\right)A^0 = \partial^0 \partial_i A^i  \,,
\end{equation}
where we have defined the effective mass
\begin{equation}
    m^2 _{A,{\rm eff}}=m^2 _A - g^2 v h + \frac{1}{2}g^2 h^2 = m^2 _A \left( 1 - \frac{h}{v} + \frac{h^2}{2v^2}\right)\,.
\end{equation}
The relativistic Lagrangian for the vector field model is then given by (see e.g. \cite{Kolb:2020fwh})
\begin{align}\label{eq:physical_Ldens_s1}
    \mathcal{L}_{A} =\frac{1}{2}\dot{A}_i\,\mathcal{P}_{ij}\,\dot{A}_j - \frac{1}{2}A_i\left(-\nabla^2 + m^2 _{A,{\rm eff}}\right)\,\mathcal{P}_{ij}\,A_j,
\end{align}
where we neglect gravitational terms for simplicity, and $\mathcal{P}_{ij}$ is the projection operator 
\begin{align}\label{eq:projection_operator}
    \mathcal{P}_{ij} = \delta_{ij} + \frac{\partial_i\partial_j}{-\nabla^2 + m_{A,{\rm eff}}^2}\,.
\end{align}
Expanding in $k^2/m^2 \ll1$ (relevant to the non-relativistic limit) and $h/v \ll 1$, the projection operator takes the form
\begin{equation}
    \mathcal{P}_{ij} \simeq \delta_{ij} + \frac{\partial_i \partial_j}{m^2 _A} \left( 1+\frac{\nabla^2}{m^2}+\frac{h}{v}+\frac{h^2}{2v^2}\right)\,.
\end{equation}
The second term appearing in the above is suppressed in the non-relativistic limit by $\partial_i \partial_j/m^2 \sim k^2/m^2$, and the corrections due to the Higgs are then doubly suppressed by $(k^2/m^2) (h/v)$ and $(k^2/m^2) (h/v)^2$. At the level of the action, the interactions generated by the corrections to $P_{ij}$ correspond to dimension-5 and dimension-6 operators, respectively. Therefore, in what follows, these terms are neglected, and the action for the vector takes the simple form,
\begin{equation}
     \mathcal{L}_A =\frac{1}{2}(\dot{A}_i)^2 - \frac{1}{2} A_i\left(-\nabla^2 + m^2 _{A,{\rm eff}}\right)A_i\,,
\end{equation}
where the interactions with the Higgs are contained entirely within the effective mass $m^2 _{A,{\rm eff}}$. The action for the physical degrees of freedom then finally takes the form

\begin{equation}\label{eq:physical-action}
\begin{split}
    S=&\int \mathrm{d}^4 x \sqrt{-g}\left[\right. \frac{1}{2}(\dot{A}_i)^2 - \frac{1}{2} A_i\left(-\nabla^2 + m^2 _{A,{\rm eff}}\right)A_i\\
    &
    +\frac{1}{2}  (\partial_{\mu }h)^2-\frac{\mh^2}{2} h^2 -\frac{\lambda}{4!}h^4 +\frac{R}{16 \pi G}\left.\right] \,.
\end{split}
\end{equation}

\subsection{Non-Relativistic Limit and the Schrodinger Equation}
 Since we are interested in dark matter and structure formation, we are interested in the non-relativistic limit of the model.  We decompose the real fields into slowly oscillating complex wavefunctions: 
\begin{align}
    \vec{A}& =\frac{1}{\sqrt{2\ma}}\left( \vec{\A}e^{-i \ma t}+{\rm c.c.}\right) \,, \\
    h& =\frac{1}{\sqrt{2\mh}}\left( \h e^{-i \mh t}+{\rm c.c.}\right).
\end{align}
We assume the wave envelope evolves on timescales short compared to $1/H$, and work in the weak gravity limit where the metric is given by the perturbed FRW metric
\begin{equation}
    ds^2 = -(1+2\Phi)dt^2+(1-2\Phi)d\bx^2,
\end{equation}
where $\Phi$ is the Newtonian potential. 
From here, we take the non-relativistic limit according to the usual procedure (laid out in \cite{Davidson:2016uok,Namjoo:2017nia,Salehian:2021khb,Ferreira_2021,Hui:2021tkt}, for instance). Terms in the action containing bare factors of $\exp( \pm imt)$ are rapidly oscillating and average to zero over times $t\gtrsim1/m$. Even so, the interaction terms must be treated with care, since they introduce a new time-scale related to the mass-splitting. For instance, the cubic interaction $hA^2$ contains terms proportional to $\exp(\pm i(\mh-2\ma))$, and violates individual conservation of particle number when $\mh=2\ma$  \footnote{For simplicity, we also avoid the case of completely degenerate masses $\mh=\ma$, which introduces resonant conversion between species.}. 
We will discuss the limits of the non-relativistic EFT momentarily. For now, we assume a safe mass hierarchy such that these terms do vanish. With this in mind, one finds
\begin{align}
    (\partial_0h)^2 &\rightarrow i(\dot{\h}\h^*-\dot{\h}^{*}\h)+\mh|\h|^2\\
    (\partial_ih)^2&\rightarrow\frac{1}{\mh}|\partial_i\h|^2\\
    h^2&\rightarrow\frac{1}{\mh}|\h|^2\\
    h^4&\rightarrow\frac{3}{2\mh^2}|\h|^4\\
    hA_i^2&\rightarrow0 \,,
\end{align}
and likewise for $A_i$. Upon substitution into Eq.~\eqref{eq:physical-action}, the non-relativistic action is finally given as
\begin{equation}\label{eq:non-rel-action}
    \begin{split}
        S=\int d^3x dt &\Big[  \frac{i}{2}\left(\vec{\A}^{\dagger}\dot{\vec{\A}}-\mathrm{c.c.}\right)- \frac{1}{2\ma}|\nabla \vec{\A}|^2-\ma\Phi|\A|^2  \\
         &+\frac{i}{2}\left(\h^*\dot{\h}-\mathrm{c.c.}\right)- \frac{1}{2\mh }|\nabla \h|^2-\mh \Phi|\h|^2  \\
         & -\frac{\lambda}{8 \mh ^2}|\h|^4+\frac{g^2}{4\ma \mh }|\A|^2|\h|^2+\frac{1}{8\pi G}\Phi \nabla^2\Phi \Big].
    \end{split}
\end{equation}
The action is invariant under the following rotations:
\begin{align}
    \A_i & \rightarrow e^{i\beta_{ij}}\A_j \\
    \h & \rightarrow e^{i\gamma}\h \,,
\end{align}
with $\gamma$ and $\beta_{ij}$ being global rotation parameters. This results in a conserved current for each degree of freedom:
\begin{align}
    \vec{j}_{A_i} &= \frac{i}{2\ma}\left( \A_i \ \nabla \A^{\dagger}_i-{\rm c.c.}\right) \,, \\
    \vec{j}_{h}&=\frac{i}{2\mh}\left(\h\nabla\h^*-{\rm c.c.}\right)\,,
\end{align}
meaning that the particle number of each species is separately conserved. 

The equations of motion are given by
\begin{eqnarray}
\label{eq:Schrodinger}
    i\partial_t \vec{\A} && = -\frac{1}{2 \ma}\nabla^2 \vec{\A}+\ma \Phi \vec{\A}- g^2\frac{\vec{\A}}{2 \ma}\frac{|\h|^2}{\mh} \,, \\
    i\partial_t \h && = -\frac{1}{2 \mh}\nabla^2 \h+\mh \Phi \h + \frac{\lambda}{2 \mh^2}\h |\h|^2- g^2\frac{\h}{2 \mh}\frac{|\vec{\A}|^2}{\ma} \,, \nonumber \\
    \nabla^2 \Phi  && = 4\pi G \left(\ma |\A|^2+\mh |\h|^2\right) \,, \nonumber
\end{eqnarray}
corresponding to a Schr\"{o}dinger equation for each wavefunction and the Poisson equation for the gravitational potential.

The structure of these equations is made manifest by rescaling to dimensionless variables, as
\begin{eqnarray}
    &&r=r_0 \hat{r}=\frac{1}{\mh ^2}\sqrt{\frac{\lambda}{G}}\hat{r}\,, \quad t=2\mh r_0^2\hat{t}\,, \\
    && \A, \h = \left(8\pi G\mh^3r_0^4\right)^{-1/2}\hat{\A},\hat{\h}\,, \quad \Phi=\frac{1}{2\mh^2r_0^2} \hat{\Phi} \,,
\end{eqnarray}
where the dimensionless quantities are denoted by \,$\hat{ }\,$.  The equations of motion then take the form, 
\begin{eqnarray}
\label{eq:dimless}
    i \hat{\A} ' &&= - \alpha\hat{\nabla}^2\hat{\A} +\frac{1}{\alpha}\hat{\Phi}\hat{\A}- \frac{1}{4\pi \alpha}\hat{\A}|\hat{\h}|^2\,, \\
    i \hat{\h} ' &&= -\hat{\nabla}^2\hat{\h} +\hat{\Phi}\hat{\h}+\frac{1}{8\pi}\hat{\h}|\hat{\h}^2|- \frac{1}{4\pi \alpha} \hat{\h}|\hat{\A}|^2 \,,\\
    \hat{\nabla}^2\hat{\Phi} &&=\frac{1}{\alpha}|\hat{\A}|^2+|\hat{\h}|^2\,,
\end{eqnarray}
where ${}'$ denotes a dimensionless time derivative, and we have defined the coupling constant
\begin{equation}\label{eq:alpha}
    \alpha \equiv \frac{\mh}{\ma}=\sqrt{\frac{2\lambda }{ g^2}}.
\end{equation}
Notably, $\alpha$ is the sole remaining free parameter of the system of equations. 

Let us now return to the validity of the non-relativistic EFT as related to $\alpha$. We note that both the regimes of heavy Higgs  ($m_h \gg m_A$) and light Higgs ($ m_h \ll m_A$) are dangerous for the EFT, the former due to decays of the Higgs to two photons of momentum $k\sim m_h \gg m_A$, and the latter due to $2\leftrightarrow2$ scattering of photons, $AA\leftrightarrow AA$ with a mediating Higgs of momentum $k\sim 2 m_A \gg m_h$. Both of these would break the non-relativistic EFT (relativistic corrections to the single-field EFT are treated systematically in \cite{Namjoo:2017nia,Salehian:2020bon,Salehian:2021khb}). Given this, in this work we restrict to $\alpha={\cal O}(1)$. Concretely, we consider the range $\alpha=[1/2,2]$.

The Schrodinger equation Eqs.~\eqref{eq:Schrodinger} can be compared with that of interacting scalar fields, such as in \cite{Mirasola:2024pmw}, with
\begin{equation}
\begin{aligned}\label{eq:interacting-schrodinger}
i  \partial _t\psi_j= & -\frac{1}{2 m_j} \nabla^2 \psi_j+m_j \Phi \psi_j \\
& +\frac{1}{2 m_j^2} \lambda_{j j}\left|\psi_j\right|^2 \psi_j+\frac{1}{4 m_j^2} \sum_k \lambda_{j k}\left|\psi_k\right|^2 \psi_j\,, \\
\nabla^2 \Phi=&4 \pi G \sum_j m_j\left|\psi_j\right|^2,
\end{aligned}    
\end{equation}
where the coupling constants $\lambda_{ij}$ can be attractive or repulsive. Comparing Eqs.~\eqref{eq:interacting-schrodinger} above to the Abelian-Higgs case in Eqs.~\eqref{eq:Schrodinger}, one may appreciate that the latter has a repulsive self-interaction for the Higgs,  an attractive interaction between the Higgs and gauge field, and the gauge field has no self-interaction. Thus one may identify the Abelian-Higgs model has a set of scalar 4-fields ($i,j=1,2,3,4$), with $i=1$ the Higgs and $i=2,3,4$ the gauge field, and $\lambda_{11}<0$ and $\lambda_{22}=\lambda_{33}=\lambda_{44}=0$, and $\lambda_{1j}>0$ for $j=2,3,4$, and $\lambda_{ij}=0$ for $i,j=2,3,4$.

\subsection{Fluid Formulation}

Finally, to lend a physical intuition to the Fuzzy Dark Sector, we may re-express the equations of motion in the language of fluid mechanics.
Making the usual Madelung decomposition (see e.g. \cite{Eberhardt:2025caq} for review),
\begin{align}
    \A_i & = \sqrt{\frac{{\rhoa}_i}{\ma}}e^{i\theta_{A_i}}\,, \\
    \h & = \sqrt{\frac{\rhoh}{\mh}}e^{i\theta_{h}} \,,
\end{align}
we replace these definitions into the equations of motion and collect real and imaginary parts. From here, let us assume $\A$ is entirely polarized in a single direction, and suppress the $\A_i$ notation. In full generality, there will be an extra copy of each equation for each component in $\A$.

The continuity equations come from the imaginary terms in the equations of motion. We find,
\begin{align}
    \dot{\rhoa}+\nabla \cdot  \vec{j}_{A} & = 0 \,,\\
    \dot{\rhoh}+\nabla \cdot \vec{j}_{h} & = 0.
\end{align}
consistent with the conservation of particle number of each species separately. 

The interactions generate new pressure terms that correspond to momentum exchange, with the Euler equations reading
\begin{align}
    \frac{D \vec{v}_{A}}{Dt}&=\frac{1}{2\ma^2}\nabla \left( \frac{\nabla^2\sqrt{\rhoa}}{\sqrt{\rhoa}}\right)-\nabla \Phi+\frac{1}{\rhoa}\nabla P_{{\rm int},h} \,, \label{eq:Euler_1} \\
    \frac{D \vec{v}_{h}}{Dt}&=\frac{1}{2\mh^2}\nabla \left( \frac{\nabla^2\sqrt{\rhoh}}{\sqrt{\rhoh}}\right)-\nabla \Phi-\frac{1}{\rhoh}\nabla P_{\rm SI}+\frac{1}{\rhoh}\nabla P_{{\rm int},A}\,, \label{eq:Euler_2} 
\end{align}
where $\frac{D}{Dt}\equiv \partial_t+\vec{v}\cdot \nabla$ is the usual convective derivative, we have introduced the bulk velocities of the form $\vec{v}=\vec{j}/\rho$, and where 
\begin{equation}
    P_{{\rm int},X}=\frac{g^2}{4\ma^2\mh^2}\rho_{X}^2\, , \quad 
    P_{\rm SI}= \frac{\lambda}{4\mh^4}\rhoh^2\, ,
\end{equation}
are the pressure for the species $X\in\{\vec{\A}, \h
\}$, which implies:
\begin{equation}
    \csh^2=\frac{\partial( P_{\rm SI}+P_{{\rm int},A})}{\partial\rhoh}=\frac{\lambda
    \rhoh}{2\mh^4}\,, \quad c_{s,A}^2=\frac{\partial P_{{\rm int},h}}{\partial\rhoa}=0.
\end{equation}
The global $U(1)$ of each field component has given unsourced continuity equations, meaning the particle number of each species is conserved. The interactions, then, offer a mutual pressure between fluid components. 

Besides the new pressure terms from self-interaction and from inter-component couplings, the model also includes, for each component, a quantum-pressure (Madelung) term, the first term on the right-hand side of Eqs. (\ref{eq:Euler_1})-(\ref{eq:Euler_2}). Present in all ULDM formulations, this term opposes gravity and contributes to the characteristic finite Jeans (characteristic) length, a common property of ULDM models.

\section{Cosmology}\label{sec:jeans}

In this work, we assume the Fuzzy Dark Sector comprises the observed cold dark matter (or a fraction thereof) and remain agnostic to the primordial production mechanism. We return to the production mechanism, and propose a minimal scenario, in Sec.~\ref{sec:production}. 

Here we study the subsequent cosmology, and in particular the linear matter power spectrum, which determines the large-scale structure of the universe in the model. 

Our approach will be to compute the Jeans length for density fluctuations in the Fuzzy Dark Sector. Following \cite{Guth:2014hsa, Ferreira:2020fam}, we do so directly from the Schrodinger-Poisson equations, Eq.~\eqref{eq:Schrodinger}. From the relation between field and density fluctuations,
\begin{eqnarray}
\rho(\bk,t) &=&  m \int  \frac{d^n k'}{(2\pi)^n} \psi^*(\bk'-\bk,t) \psi(\bk',t),
\label{eq:densitymode}
\end{eqnarray}
we infer that an instability for $\psi$ on a scale $k_*$ implies an instability for $\rho$ on a scale $k_J = k_*$.

Now, let us focus on the field fluctuations. Consider a homogeneous and isotropic background condensate in both fields, with small perturbations
\begin{align}
    \A(\bx,t)&=\A_c(t)+\delta \A(\bx,t)\,, \\
    \h(\bx,t)& =\h_c(t)+\delta \h(\bx,t). 
\end{align}
We assume for simplicity that $\vec{\A}$ is linearly polarized. The zeroth-order equations admit condensate solutions
\begin{align}
    \A_c &= \A_0 e^{-i \mu_{A}t} \,,\\
    \h_c &= \h_0 e^{-i \mu_{h}t} \,,
\end{align}
with chemical potentials
\begin{align}
    \mu_{A} &= -\frac{g^2}{2\ma\mh}\h_0^2 \,, \\
    \mu_{h} & = \frac{\lambda}{2\mh^2}\h_0^2 - \frac{g^2}{2\ma \mh}\A_0^2 .
\end{align}
We then decompose the perturbations into real and imaginary parts
\begin{align}
    \frac{\delta\A}{\A_c}&=A+iB \,, \\
     \frac{\delta\h}{\h_c}&=C+iD \,,
\end{align}
leading to the system of equations in Fourier space:
\begin{widetext}
\begin{equation}\label{eq:matrix}
    \partial_t
    \begin{pmatrix}
        A \\
        B\\
        C\\
        D
    \end{pmatrix}
     = 
     \begin{pmatrix}
         0 & \frac{k^2}{2\ma } & 0 & 0 \\
         -\frac{k^2}{2\ma }+\frac{8\pi G \ma ^2}{k^2}|\A_c|^2 & 0 &  \left(\frac{g^2}{\ma  \mh}+\frac{8\pi G \ma  \mh}{k^2}\right)|\h_c|^2 & 0 \\
         0 & 0 & 0 & \frac{k^2}{2\mh^2} \\
         \left(\frac{g^2}{\ma  \mh}+\frac{8\pi G \ma  \mh}{k^2}\right)|\A_c|^2 & 0 & -\frac{k^2}{2\mh}-\frac{\lambda}{\mh^2}|\h_c|^2+\frac{8 \pi G \mh^2}{k^2}|\h_c|^2 & 0
     \end{pmatrix}
     \begin{pmatrix}
         A \\
         B \\
         C \\
         D
     \end{pmatrix} \,,
\end{equation}
with the corresponding secular equation
\begin{equation}
\begin{split}
    &\omega^2=2 \pi 
   G \left( \ma  \na+\mh \nh\right)-\frac{k^4}{8 \ma ^2}-\frac{k^4}{8 \mh^2}-\frac{\lambda k^2 \nh}{4 \mh^3} \pm \frac{1}{2}\Bigg[\left(-4 \pi  G \ma  \na+\frac{k^4}{4 \ma ^2}-4 \pi  G \mh \nh+\frac{k^4}{4 \mh^2}+\frac{\lambda k^2 \nh}{2 \mh^3}\right)^2\\&
   +\frac{16 \pi  g^2 G k^2 \nh \na}{\ma  \mh}+\frac{4\pi  G k^4 \mh \nh}{\ma ^2}+\frac{4\pi  G k^4 \ma 
   \na}{\mh^2}+\frac{8 \pi  G \lambda k^2 \ma  \nh \na}{\mh^3}+\frac{g^4 k^4 \nh \na}{ \ma ^3 \mh^3}-\frac{k^8}{4 \ma ^2 \mh^2}-\frac{\lambda k^6 \nh}{2 \ma ^2 \mh^3}\Bigg]^{1/2},
   \end{split}
\end{equation}
\end{widetext}
having introduced number densities $n_h=|\h_c|^2$ and $n_A=|\A_c|^2$.

Perturbations will grow if at least one solution for $\omega$ is imaginary. The associated $k_*$ below which $\omega^2<0$, leading to growing solutions, is
\begin{eqnarray}
k_*^2=&&\frac{\kJh^4+\kJa^4+4 k_g^4+\frac{4}{3} \mh^4\csh^4}{f(k_g,\csh,k_{G,X})}-\frac{2}{3} \mh^2\csh^2 \nonumber \\
&& +\frac{1}{3} F(k_g,\csh,k_{G,X})\label{eq:k*}  \,,
\end{eqnarray}
where we have introduced the following
\begin{align}
    &k_{G,X}^4=16\pi Gm_X^3n_X,\\
    &k_g^4=\frac{g^4\nh\na}{\ma  \mh}, \\
    &\csh^2=\frac{\lambda \nh}{2 \mh^3},
\end{align}
where $k_{G,X}$ is the usual gravitational Jeans length, and the function $F$ is defined as
\begin{widetext}
\begin{equation}
\begin{split}
    &F^3(k_g,\csh,k_{G,X})=-18 \mh^2\csh^2 \left(\kJh^4-2 \kJa^4+4 k_g^4\right) +\frac{27 k_g^4 \kJh^4}{2\mh^2\csh^2}-64 \mh^6\csh^6
    \\
    &+\left(\frac{4 \left(9 k_g^4 \left(16 \mh^4\csh^4-3 \kJh^4\right)+36 \mh^4\csh^4
   \left(\kJh^4-2 \kJa^4\right)+128 \mh^8\csh^8\right)^2}{16\mh^4\csh^4}\right.
   \left.- \left(3 \left(\kJh^4+\kJa^4\right)+12 k_g^4+16 \mh^4\csh^4\right){}^3  \vphantom{\frac{\left(\mh^4\right)}{\mh^4}} \right)^{1/2}.
\end{split}
\end{equation}
\end{widetext}
In the limit of vanishing $n_{h,A}$, then, we have
\begin{align}
    &\lim_{n_h\rightarrow0}k_*^2=k_{G,A}^2\label{eq:kstar-lim1} \,,\\
    &\lim_{n_A\rightarrow0}k_*^2=\sqrt{k_{G,h}^4+4m_h^4c_{s,h}^4}-2m_h^2c^2_{s,h} \,.\label{eq:kstar-lim2}
\end{align}
This $k_*$ defines the characteristic scale of the FDS, in analogy to the Jeans length $k_J$ of the single-field fuzzy dark matter and cold dark matter.

Remarkably, there is a {\it single} characteristic scale for the growth of perturbations in the multicomponent Fuzzy Dark Sector. Intuitively, this arises due to the intrinsically interacting nature of the system: The mass of the dark photon is itself generated by interactions with the Higgs.  

With the characteristic scale in hand, we can now turn to observables. Since the model features a single scale, the imprint on the CMB, LSS, and other cosmological probes is simply the single-field Fuzzy Dark Matter with the modification that the Jeans length $k_J$ be replaced by $k_*$ defined in Eq.~\eqref{eq:k*}.  We can write the linear matter power spectrum at $z=0$ as
\begin{equation}
    P_{\rm FDS}(k) = T_{\rm F} ^2 (k) P_{\rm CDM}(k) \,,
\end{equation}
where the transfer function is given by \cite{PhysRevLett.85.1158}
\begin{equation}
    T_{\rm  F}(k) \simeq \frac{\cos(x ^3)}{1+ x^8} \,,
\end{equation}
where
\begin{equation}
    x = 1.3 (k_* \, {\rm Mpc})^{1/9} \frac{k}{k_{*,{\rm eq}}} \,,
\end{equation}
with $k_{*,{\rm eq}}=a_{\rm eq}^{1/4}k_*$ \cite{Hu_2000} and where $k_*$ encodes the FDS modifications to the usual physics of single-field fuzzy dark matter.

\begin{figure*}
    \centering
    \includegraphics[width=.7\linewidth]{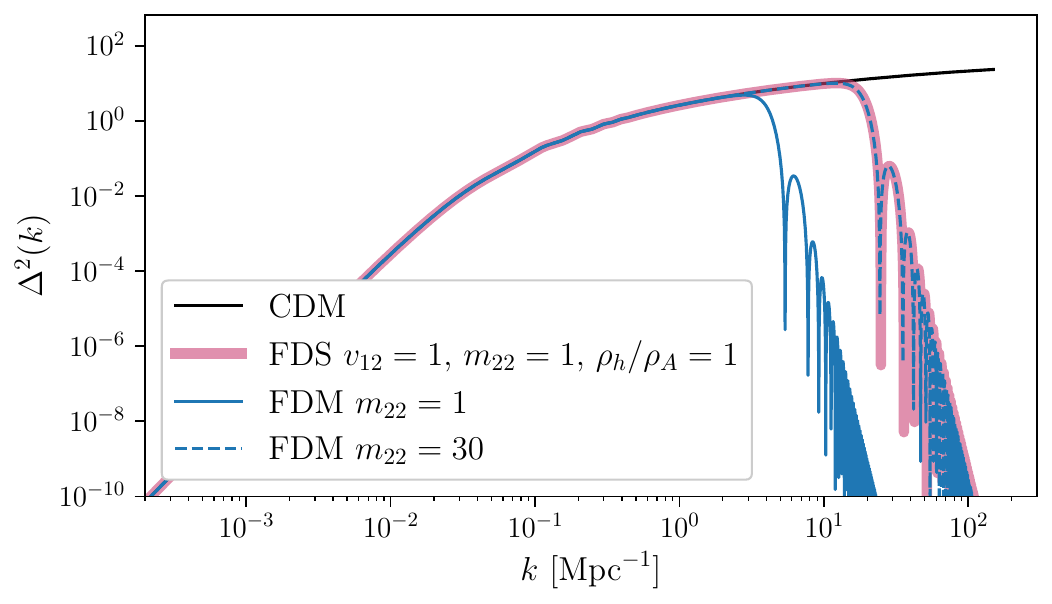}
    \caption{Matter power spectrum in the Fuzzy Dark Sector (FDS). Shown for comparison are FDM and CDM. Here we have chosen reference scales $v_{12}=v/(10^{12} \ {\rm GeV})$ and $m_{22}=m/(10^{-22} \ {\rm eV})$. For $v_{12}\leq 1$ we find the FDS mimics a heavy FDM candidate, whereas for $v_{12}\geq 1$ it mimics a lighter FDM candidate.}
    \label{fig:Pk}
\end{figure*}

From this, one can easily construct the linear matter power spectrum in the fuzzy dark sector. The result is shown in Fig.~\ref{fig:Pk}, where we assume a fiducial cosmology with the Planck 2018 best-fit $\Lambda$CDM parameters \cite{Planck:2018vyg}. One may appreciate that the cut in the matter power spectrum appears at a larger value of $k$ than in FDM with the same mass, thereby mimicking a heavier FDM candidate for an appropriate choice of parameters. The sign of this correction (shifting the cut to higher or lower $k$) can be more easily appreciated by considering the small-$\lambda$ expansion of Eq.~\eqref{eq:k*}:
\begin{equation}
k_*^2\approx\sqrt{\kJh^4+\kJa^4}+2\mh^2\csh^2\left(4-\frac{ 5}{1+\frac{\kJa^4}{\kJh^4}}\right),
\end{equation}
where the cut will be shifted to higher $k$ if $\kJa^4/\kJh^4\gtrsim\frac{1}{4}$.   

\section{Solitons}\label{sec:solitons}

We now establish a {\it diversity} of soliton profiles, corresponding to the central cores of dark halos in the Fuzzy Dark Sector model. This is analogous to the two-field fuzzy dark matter presented in e.g. \cite{Luu:2024lfq}.

Following \cite{Schiappacasse:2017ham}, the Hamiltonian can then be broken into kinetic, interacting, and gravitational terms:
\begin{equation}
    H=H_{\rm kin}+H_{\rm int}+H_{\rm grav},
\end{equation}
where we have dropped the Einstein-Hilbert term, and
\begin{align}
    H_{\rm kin} & = \int d^3x \left( \frac{1}{2\ma }|\nabla\A|^2+\frac{1}{2\mh}|\nabla\h|^2 \right) \,,\\
    H_{\rm int} & = \int d^3x \left(\frac{\lambda}{8 \mh^2}|\h|^4-\frac{g^2}{4\ma \mh}|\A|^2|\h|^2 \right) \,, \\
    H_{\rm grav} & = \int d^3x  \left( \Phi \left(\ma |\A|^2+\mh|\h|^2 \right)\right).
\end{align}
In order to estimate basic soliton properties, let us take the ansatz
\begin{align}
    \vec{\A} &= \vec{\epsilon} \ \sqrt{\frac{M_{A}}{\pi^{3/2} \ma  R_A^3}}e^{-i \mu_A t}e^{-\frac{r^2}{2R_A^2}} \,,\\
    \h &=  \sqrt{\frac{M_{h}}{\pi^{3/2}  \mh R_h^3}}e^{-i \mu_h t}e^{-\frac{r^2}{2R_h^2}}  \,,
\end{align}
with total particle number $N_X=\int d^3x |X|^2$, chemical potential $\mu_X$, and radius $R_X$ for the species $X\in\{\vec{\A}, \h \}$. For simplicity, we assume that the vector soliton is entirely linearly polarized along the unit $\vec{\epsilon}$ direction, though additional polarization states can lead to additional interesting soliton phenomenology \cite{Zhang:2021xxa,Gorghetto:2022sue,Jain:2021pnk,Amin:2022pzv,Wang:2023tly}. 
The gravitational potential is exactly solvable, with
\begin{equation}
\Phi(r)=-\frac{G}{r}\left[ \mh N_h {\rm erf}\left(\frac{r}{R_h}\right)+\ma N_A {\rm erf}\left(\frac{r}{R_A}\right) \right] \,,
\end{equation}
where we can see in the far-field limit that this recovers the expected potential for two point sources:
\begin{equation}
    \Phi \sim -\frac{G }{r}\left( \ma N_A+\mh N_h \right).
\end{equation}
The variational procedure is as follows: we insert our simplistic ansatz into the full Hamiltonian, and then, leveraging the analytical expression for the gravitational potential $\Phi$, vary the resulting function $H(N_X,R_X)$ in order to find the radius parameters ${R_X}$ that minimize the Hamiltonian.

Integrating the Gaussian ansatz, the constituent pieces of the Hamiltonian become:
\begin{align}
    H_{\rm kin} & = \frac{3M_h}{4\mh^2
   R_h^2}+\frac{3M_A}{4\ma^2 R_A^2} \,,\\
    H_{\rm int} & = \frac{\sqrt{2}\lambda  M_h^2}{32 \pi^{3/2}  \mh^4 R_h^3}-\frac{g^2 M_h
   M_A}{4 \pi^{3/2}  \mh^2 \ma^2 (R_h^2+R_A^2)^{3/2}} \,,\\
    H_{\rm grav}&=-\frac{G}{\sqrt{\pi}} \left[\frac{\sqrt{2}M_h^2}{R_h}+\frac{
   \sqrt{2}M_A^2}{R_A}+\frac{4 M_h M_A
   }{(R_h^2+R_A^2)^{1/2}}\right].
\end{align}
What remains is to extremize $H$ with respect to $R_A, R_h$. The criterion for stable solitons is that the local extrema in $H$ have a positive Hessian matrix.

\begin{figure*}    \includegraphics[width=1\linewidth]{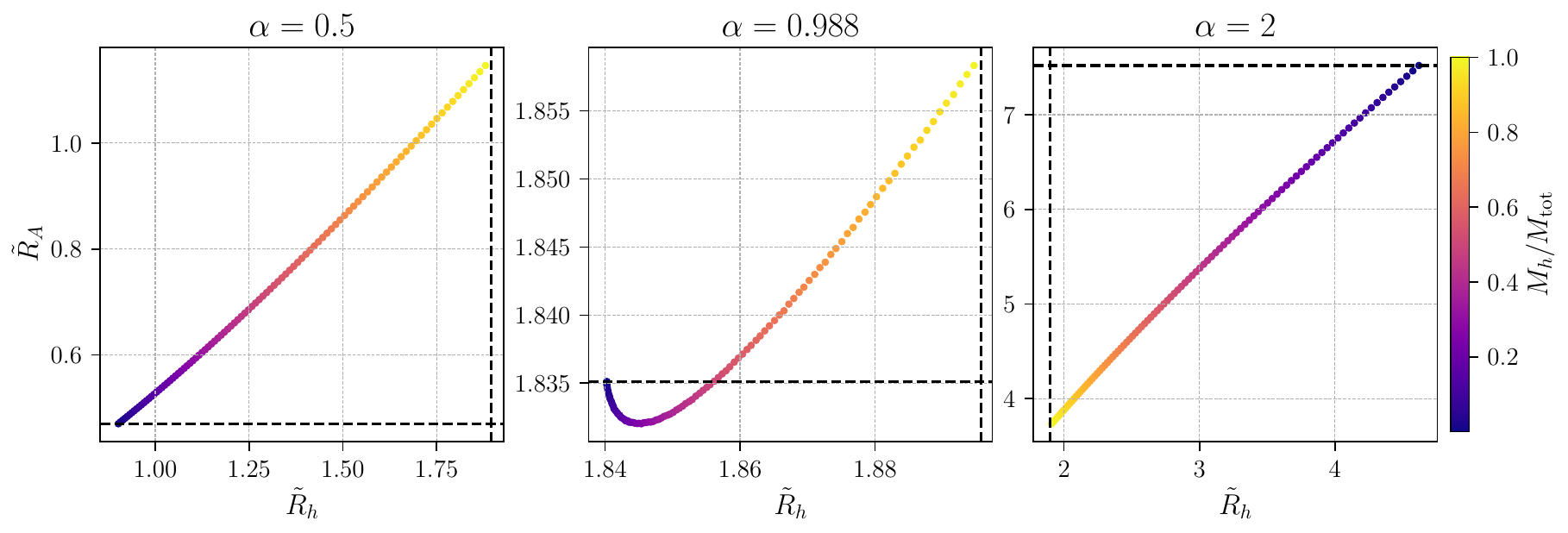} 
    \caption{Solitons of fixed total mass. Soliton radii $\tilde{R}_{A,h}$ which extremize the Hamiltonian Eq.~\eqref{eq:dimensionless-ham} for particle mass ratio $\alpha=0.5$ (left), $\alpha=0.988$ (center), and $\alpha=2$ (right). The total mass is fixed at $\tilde{M}_{\rm tot}=1$ across each plot, with the color gradient indicating the relative mass fraction in each species. Horizontal and vertical dashed lines correspond to the radii in Eq.~\eqref{eq:Ra-limit} and \eqref{eq:Rh-limit} respectively. We show $\alpha=0.988$ as this is where $\tilde{R}_A(\tilde{M}_h=0)=\tilde{R}_A(\tilde{M}_h=0.5)$ numerically. See the main text for discussion.}
\label{fig:parameter-scan}
\end{figure*}

Let us re-cast our parameters in terms of the following dimensionless quantities:
\begin{align}
    H &=H_0\tilde{H} = \mh^2\sqrt{\frac{G}{\lambda^3}}\tilde{H} \,,\\
    R_X  &= r_0\tilde{R}_{X} =\frac{1}{\mh^2}\sqrt{\frac{\lambda}{G}}\tilde{R}_X \,,\\
    M_X & = m_0\tilde{M}_{X}=\frac{1}{\sqrt{\lambda G}}\tilde{M}_X \,,
\end{align}
having fixed physical scales $H_0$, $r_0$, and $m_0$ in terms of $\mh$, $\lambda$, and $G$. Hence, using $\lambda=\mh^2/2v^2$, we have
\begin{align}
    R_X&=552 \ {\rm kpc}\left(\frac{10^{12}\ {\rm GeV}}{v}\right)\left(\frac{10^{-22} \ {\rm eV}}{\mh}\right)\tilde{R}_X \,,\\
    M_X&=1.55\times10^5 \ {M_{\odot}}\left(\frac{v}{10^{12}\ {\rm GeV}}\right)\left(\frac{10^{-22} \ {\rm eV}}{\mh}\right)\tilde{M}_X.
\end{align}
With these rescalings, the dimensionless Hamiltonian may be read off as
\begin{equation}\label{eq:dimensionless-ham}
    \begin{split}
    \tilde{H}= & \frac{3\tilde{M}_h}{4
   \tilde{R}_h^2}+\frac{3\alpha^2\tilde{M}_A}{4\tilde{R}_A^2}+\frac{\sqrt{2}  \tilde{M}_h^2}{32 \pi^{3/2}  \tilde{R}_h^3}-\frac{\alpha \tilde{M}_h
   \tilde{M}_A}{4 \pi^{3/2} (\tilde{R}_h^2+\tilde{R}_A^2)^{3/2}} \\
        &-\frac{\sqrt{2}\tilde{M}_h^2}{\sqrt{\pi}\tilde{R}_h}-\frac{
   \sqrt{2}\tilde{M}_A^2}{\sqrt{\pi}\tilde{R}_A}-\frac{4 \tilde{M}_h \tilde{M}_A
   }{\sqrt{\pi}(\tilde{R}_h^2+\tilde{R}_A^2)^{1/2}}.
    \end{split}
\end{equation}
We minimize this Hamiltonian numerically, scanning over the ratio of particle masses $\alpha$ defined in Eq.~\eqref{eq:alpha}, with solutions shown in Fig.~\ref{fig:parameter-scan}.

In the limit of vanishing $M_{h,A}$, the soliton radii are
\begin{align}    &\lim_{M_h\rightarrow0}\tilde{R}_A=\frac{3}{2}\sqrt{\frac{\pi}{2}}\frac{\alpha^2}{\tilde{M}_{\rm tot}} \,, \label{eq:Ra-limit}\\
    &\lim_{M_A\rightarrow0}\tilde{R}_h=\frac{1}{4 \tilde{M}_{\rm {tot}}}\left( 3 \sqrt{\frac{\pi }{2}}+\sqrt{\frac{3}{2 \pi }} \sqrt{\tilde{M}_{\rm {tot}}^2+3 \pi ^2}\right)\label{eq:Rh-limit}.
\end{align}
Notice that because the Higgs has repulsive self-interaction, Eq.~\eqref{eq:Rh-limit} features a relative sign difference compared to the ALP case (for instance Eq. (41) of \cite{Schiappacasse:2017ham}). As such, $\tilde{R}_h$ is strictly real, and hence stable for all $\tilde{M}_{\rm tot}$. Likewise for $\tilde{R}_A$, which lacks any self-interactions. 

The connection to the Jeans analysis can be made more explicit upon the identification $R\sim1/k$ and $M\sim\rho/k^3$. Restoring physical units, the variational method gives

\begin{align}
    \lim_{\rho_h\rightarrow0}k_A^2&\sim  \sqrt{\sqrt{\frac{2}{\pi}}\frac{20}{3}G\ma^2\rho _A} \,,\\
    \lim_{\rho_A\rightarrow0}k_h^2&\sim\sqrt{\sqrt{\frac{2}{\pi}}\frac{20}{3}G\mh^2\rho_h +\left(\frac{3}{48}\right)^2\frac{2}{\pi}\frac{\lambda^2 \rho_h^2}{m_h^4}}-\frac{3}{48}\sqrt\frac{2}{\pi}\frac{\lambda \rho_h}{m_h^2},
\end{align}
which is consistent with Eq.~\eqref{eq:kstar-lim1} and \eqref{eq:kstar-lim2} up to numerical prefactors. Because the dark photon has no self-interaction pressure support, the $\rho_h=0$ limit is just the usual gravitational Jeans length.

\begin{figure}
    \includegraphics[width=1\linewidth]{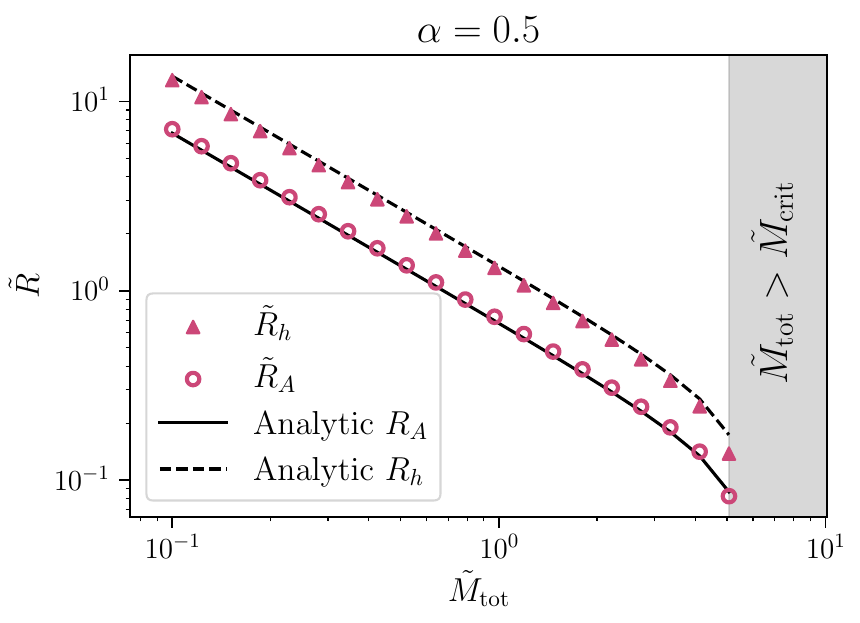}
    \caption{Dependence of $\tilde{R}_{h,A}$ on the combined soliton mass $\tilde{M}_{\rm tot}$, with $\alpha=0.5$. The mass fraction in each species is fixed at $\tilde{M}_h=\tilde{M}_A=\tilde{M}_{\rm tot}/2$. Colored markers are the numerical result using the full Hamiltonian, whereas solid lines are the analytical approximations Eq.~\eqref{eq:Rh-analytic} and \eqref{eq:Ra-analytic}, discussed in the text. The shaded region denotes where the analytic expressions become imaginary, as well as where the numerical solver fails.}
    \label{fig:total-mass}
\end{figure}

To make further analytical progress, we take $\tilde{R}_A= \alpha\tilde{R}_h$, to be justified \textit{a posteriori}. In this case,
\begin{widetext}
\begin{align}
    \tilde{R}_A&\approx \alpha\tilde{R}_h \,,\label{eq:Ra-analytic}\\
    \tilde{R}_h&\approx \frac{3 \pi \alpha \beta^2+\sqrt{3\alpha}  \sqrt{3 \pi^2 \alpha \beta^4+f\left[-32 f \bar{f}^2 \alpha^2+f\left(\bar{f}^2+\alpha f^2\right) \beta^4+2 \sqrt{2} \bar{f} \alpha \beta\left(-4 \bar{f}^2+f^2\left(\beta^2-4 \alpha\right)\right)\right] \tilde{M}_{\rm tot}^2}}{4 \sqrt{\pi} \beta\left(4 f \bar{f} \alpha+\sqrt{2}\left(\bar{f}^2+\alpha f^2\right) \beta\right) \tilde{M}_{\rm tot}} \,, \label{eq:Rh-analytic} 
\end{align}
\end{widetext}
where we have introduced $\beta^2=(1+\alpha^2)$, $f=M_h/M_{\rm tot}$, and $\bar{f}=(1-f)=M_A/M_{\rm tot}$. The above then recover Eq.~\eqref{eq:Ra-limit} and \eqref{eq:Rh-limit} in the limit $f=0$ and $\bar{f}=0$, respectively.  These expressions are compared to the full numerical result in Fig.~\ref{fig:total-mass}, showing excellent agreement and justifying the assumption $\tilde{R}_A=\alpha \tilde{R}_h$.

Despite the restricted range of $\alpha$ that we consider, interactions have a dramatic influence on soliton formation. One may appreciate from Fig.~\ref{fig:total-mass} that solitons cannot form above a critical mass $\tilde{M}_{\rm crit}$, given by
\begin{equation}
\begin{split}
    \tilde{M}_{\rm crit}\approx \pi  \sqrt{3\alpha } \beta ^2\left[-f \left(\beta ^4 f (\bar{f}^2+\alpha f^2)-32 \alpha ^2 f \bar{f}^2\right.\right.
    \\\left.\left.
    +2 \sqrt{2} \alpha  \beta  \bar{f} \left(f
   -4\bar{f}^2+f^2\left(\beta^2-4 \alpha\right)\right)-4\right)\right]^{-1/2}.
\end{split}
\end{equation}
Namely, Eq.~\eqref{eq:Rh-analytic} becomes complex for large enough total mass. This instability is a direct result of the coupling between fields. In the fully decoupled limit, taking $H_{\rm int}\rightarrow0$ by hand (or likewise $f,\bar{f}=0$), we indeed find that solutions $\tilde{R}_{A,h}$ freely extend to arbitrarily large masses.

\begin{figure}
    \centering
    \includegraphics[width=.9\linewidth]{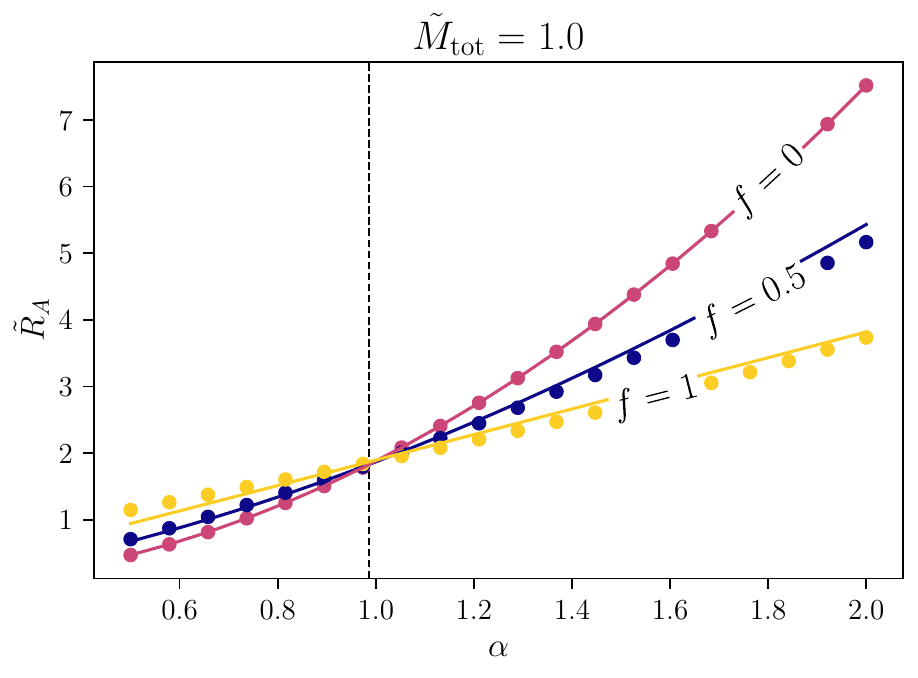}
    \caption{Vector soliton radius as a function of $\alpha$. Each curve corresponds to a different mass fraction $f=M_h/M_{\rm tot}$. Solid lines correspond to the analytical expressions Eq.~\eqref{eq:Ra-limit} and \eqref{eq:Ra-analytic} for $\tilde{M}_h\rightarrow0$ and $\tilde{M}_h\rightarrow\tilde{M}_{\rm tot}/2, \ \tilde{M}_{\rm tot}$, respectively. Colored markers represent numerical solutions. The analytical results for $f=0,0.5$ intersect at $\alpha= 1$. The crossover is found numerically to be at $\alpha=0.988$, denoted by a vertical dashed line.}
    \label{fig:crossover}
\end{figure}

To further elucidate the dependence on $\alpha$, we sample solitons finely in the $\{R,\alpha \}$ plane. This is shown in Fig.~\ref{fig:crossover}, where the analytical expressions as well as numerical solutions $\tilde{R}_A$ are shown for different reference mass fractions $f\in\{0,0.5,1\}$. We find that solutions $\tilde{R}_A(f)$ cross and hence exchange ordering near $\alpha\approx1$:

\begin{equation}
    \begin{split}
    &R_A(0)<R_A(0.5)<R_A(1) \,, \quad {\rm for} \ \alpha<1 \,,\\
    &R_A(0)>R_A(0.5)>R_A(1)\,, \quad {\rm for} \ \alpha>1\,,
    \end{split}
\end{equation}
providing a further explanation, observed turn-around in the middle panel of Fig.~\ref{fig:parameter-scan}. 

\begin{figure}
    \centering
    \includegraphics[width=1\linewidth]{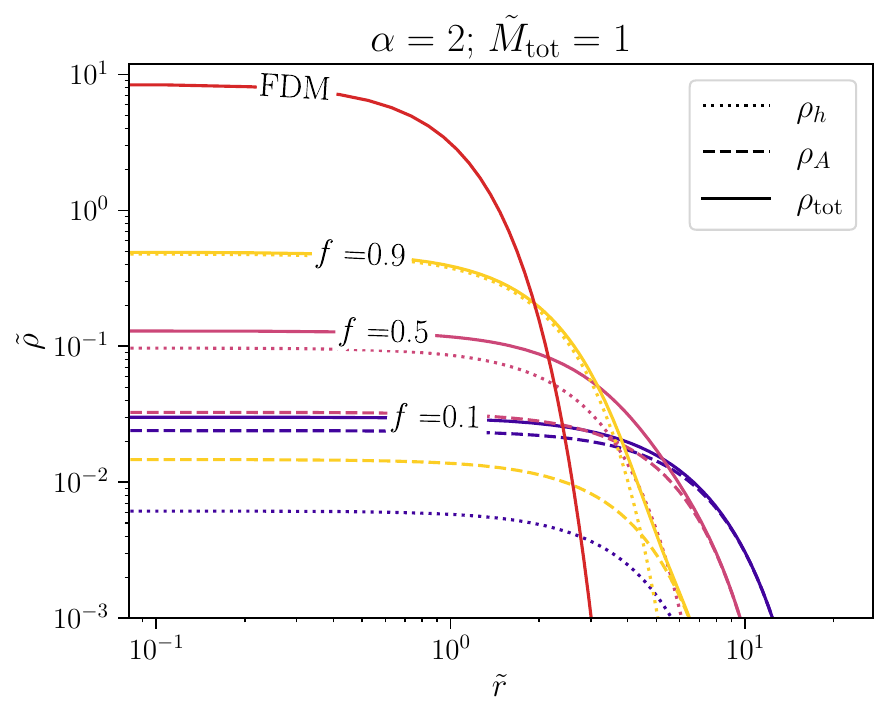}
    \caption{Diversity of density profiles coming from varying the mass fraction $f=M_{h}/M_{\rm tot}$ at fixed $\alpha=2$ and fixed total soliton mass $\tilde{M}_{\rm tot}=1$. The equal-mass FDM soliton is overlaid in red.}
    \label{fig:density-profiles}
\end{figure}

Finally, we turn our attention to the diversity of solitons in the Abelian-Higgs Fuzzy Dark Sector. This is illustrated in Fig.~\ref{fig:density-profiles}, where we show the density profile for different mass fractions, at fixed $\alpha=2$ and fixed total mass $\tilde{M}_{\rm tot}$. The colors indicate differing mass fraction, with dotted and dashed lines the Higgs and gauge field contributions, respectively. We compare against an equal mass FDM soliton with no self-interactions.

For these examples shown in Fig.~\ref{fig:density-profiles}, the central density ranges over an order of magnitude, from $0.03$ to $0.49$, and the soliton radius ranges over a factor of $\sim 3$, from 2.11 to 6.55. We find that the diversity of soliton radii increases for \textit{smaller} $M_{\rm tot}$, since $R_{A,h}\sim1/M_{\rm tot}$. However, \textit{larger} $M_{\rm tot}$ leads to a greater diversity in the central densities of solitons, since $\rho_{\rm tot}\sim M_{\rm tot}$. Thus, exploring the full space of $\{\alpha,M_{\rm tot}\}$ and restoring the extra polarization modes, we can expect a great diversity of dark halos in the Abelian-Higgs fuzzy dark sector.

\section{Discussion} \label{sec:discussion}


To close, we consider interesting and important aspects of the FDS that lie beyond the scope of the current work. In particular, we consider the primordial production mechanism for the Fuzzy Dark Sector, the portals to the Standard Model, and alternative model realizations of the FDS paradigm.

\subsection{Production Mechanism}
\label{sec:production}

In Sec.~\ref{sec:jeans}, we assumed a priori that the Fuzzy Dark Sector provides the observed relic density of dark matter or an ${\cal O}(1)$ fraction thereof. This requires a viable production mechanism consistent with the cold dark matter paradigm. While scalars and dark photons have a multitude of separate production mechanisms, such as dark photon production from parametric resonance \cite{Dror:2018pdh,Agrawal:2018vin,Adshead:2023qiw}, it remains an open problem to find a mechanism that can directly link the relic density of the scalar and vector fields. 

Here we propose a minimal model to produce both fields {\it without introducing any new fields into the model}, nor any Standard Model interactions. The proposed model has an action given by
\begin{eqnarray}
S=&\int \mathrm{d}^4 x \sqrt{-g} \left[\right.\frac{R}{16 \pi G}-\frac{1}{4} F_{\mu \nu} F^{\mu \nu}+\frac{1}{2}  |D_{\mu}\phi|^2  \\
&-\frac{\lambda}{4}\left(|\phi|^2-v^2\right)^2 + g_{\theta AA}\theta F \tilde{F} + \Lambda^4 \cos(\theta)\left. \right], \nonumber
\end{eqnarray}
where $\theta$ is the phase of the Higgs field, $\phi = (v+h)e^{i\theta}$. This model leads to parametric resonance of the Higgs $h$ via the kinetic coupling $h^2(\partial \theta)^2$ and a tachyonic instability of the gauge field due to the Chern-Simons interaction $g_{\theta A A} \theta F\tilde{F}$. By adjusting $\Lambda$ and $g_{\theta A A}$, one expects a comparable relic density can be achieved, though whether this can be achieved in a manner consistent with {\it naturalness} remains an open question.

Another production mechanism that allows for a secluded or `completely dark' FDS is that of gravitational particle production (GPP). This has been studied independently for both scalars \cite{Jenks:2024fiu} and vectors \cite{Kolb:2020fwh,Kolb:2022eyn,Capanelli:2024rlk}. In \cite{Capanelli:2024pzd}, the case of a non-minimally coupled vector in the Abelian-Higgs model was considered, although the Higgs was taken to be heavy. The simultaneous production of a light scalar and vector through GPP has not yet been considered. 

Finally, if either the dark photon or Higgs is coupled to the Standard Model, they may be \textit{frozen-in} from the SM bath. Indeed, even if only one species has SM interactions, there can be a variety non-standard freeze-in scenarios that populate the interacting dark sector \cite{Evans:2019vxr,Fernandez:2021iti,Bhattiprolu:2024dmh}. We discuss some of the possible Standard Model portals that allow for freeze-in below.

\subsection{Portals to the Standard Model}

The Abelian-Higgs Fuzzy Dark Sector can couple to the Standard Model both through the standard dark photon channels and by couplings of the dark Higgs.

For example, the Higgs could be {\it millicharged} under the SM U(1), leading to the interaction $|D_\mu \phi|^2 \supset \varepsilon h\partial_\mu h A^{\mu(\rm vis)}$, or coupled to the SM U(1) field strength $F_{\mu \nu}^{(\rm vis)}$ via the quadratic coupling $h^2 F_{\mu \nu}^{(\rm vis)}F^{\mu \nu(\rm vis)}$. The ultra-light dark matter phenomenology of these portals has been discussed in \cite{Jaeckel:2021xyo} for the millicharged case and \cite{Arvanitaki:2014faa} for the quadratic coupling.

The dark photon, on the other hand, can couple to any of the SM leptons or quarks via the interaction $\varepsilon A_\mu j^{\mu}$ where $j^{\mu}$ is a conserved current of the Standard Model. The dark photon can also kinetically mix with the SM photon, $\varepsilon F_{\mu \nu} ^{(\rm vis)} F^{\mu \nu {(\rm dark)}}$, leading to visible-dark photon oscillations and its own wealth of phenomenology \cite{An:2013yfc,Redondo:2013lna,Caputo:2020bdy,Caputo:2020rnx,An:2020jmf,Amin:2023imi,Romanenko:2023irv,Brahma:2023zcw, An:2023wij,  Brahma:2025vdr}.

As a final example, one expects both the dark Higgs and dark photon to couple generically through the SM Higgs portals of the form $h^2H^{\dagger}H$ and $g^{\mu\nu}A_{\mu}A_{\nu}H^{\dagger}H$, respectively.  Astrophysical bounds (e.g. stellar emissions) impose strong constraints on the Higgs mass and mixing angle \cite{Dicus:1978fp, Dev:2020jkh,DeRocco:2020xdt} as well as on Higgsed vector bosons \cite{An:2020bxd}. Extensions of these results to an interacting FDS may provide interesting future insight.

Even relaxing the assumption that the FDS comprises all of the dark matter, an irreducible background abundance of FDS particles can be produced through any of the above portals (similar to the case of ALPs \cite{Langhoff:2022bij} or millicharged particles \cite{Iles:2024zka}), and a host of late-time phenomenology realized. We leave the exploration of the Fuzzy Dark Sector phenomenology of these portals to the Standard Model to future work.

\subsection{Dawn of the Fuzzy Dark Sectors: More Models to realize the FDS paradigm}

In the present work we have introduced only a single realization of the FDS paradigm. However, many FDS models are possible. For example, in the context of a confining gauge theory, there can be many interacting light degrees of freedom (see e.g. \cite{Maleknejad:2022gyf,Alexander:2023wgk,Jain:2023ojg,Alexander:2024nvi}). Here we outline one minimal example, with ultra-light radial and angular modes in a complex scalar field:
\begin{equation}
\begin{split}
S[\Psi,\partial_{\mu}\Psi]=&\int d^4x\sqrt{-g}\big[ \frac{\mathcal{R}}{16\pi G} +\frac{1}{2}\partial_{\mu}\Psi\partial^{\mu}\Psi^* \\
&-\lambda(|\Psi|^2-f_a^2)^2 - \Lambda^4_a\left(1-\cos{\ a/f_a} \right)\big] \,,
\end{split}
\end{equation}
where $\Psi=(f_a+h)e^{a/f_a}$, with $h$ the dark Higgs and $a$ an ALP with the usual cosine potential.
This model is similar but distinct from \cite{Toomey:2025mvx}. Taking the non-relativistic limit, and decomposing the radial and angular fluctuations into wavefunctions $h\sim\h e^{-i\mh t}$ and $a\sim\psi e^{-im_at}$ respectively, leads to a modified Schr\"odinger equation for $\h$:
\begin{equation}\label{eq:radial_schrodinger}
i\dot\h=-\frac{\nabla^2\h}{2m_{h}}+m_{h}\Phi\h+\frac{3\lambda}{m_{h}^2}|\h|^2\h -\frac{\lambda_a}{2m_a \mh}|\psi|^2\h,
\end{equation}
and likewise for $\psi$:
\begin{equation}\label{eq:angular_schrodinger}
\begin{split}
    \big(1+\frac{4\lambda|\h|^2}{m_{h}^3} \big)i\dot\psi =& \big(1+\frac{4\lambda|\h|^2}{m_{h}^3}\big)\big(-\frac{\nabla^2\psi}{2m_a}\big)+m_a\Phi\psi  \\
    -&\frac{3\lambda_a}{m_a^2}|\psi|^2\psi+\frac{\lambda_a}{2m_a\mh }|\h|^2\psi,
\end{split}
\end{equation}
with the Poisson equation
\begin{equation}\label{eq:Poisson}
\nabla^2\Phi=4\pi G \{ m_{h}|\h|^2+m_{a}|\psi|^2\} \,,
\end{equation}
where $\mh^2=4\lambda f_a^2$, $m_a^2=\Lambda_a^4/f_a^2$, and $\lambda_a=\Lambda_a^4/4!f_a^4$.  Despite its simplicity, this Fuzzy Dark Sector already has several interesting features: (1) The VEV of the Higgs coincides with the ALP decay constant. While this choice can be relaxed so that $\langle\Psi\rangle\neq f_a$, this is degenerate with freely varying $\lambda$. 
(2) The radial and angular self-interactions carry opposite signs: one is attractive while the other is repulsive. As such, we expect a diversity of \textit{non-spherical} (e.g. hollow or separate) soliton configurations, as found in \cite{Mirasola:2024pmw}.
Finally, (3)  The model features a kinetic coupling, e.g. the non-canonical form of Eq.~\eqref{eq:angular_schrodinger}, as recently studied in \cite{Toomey:2025mvx}. While this coupling is descended from a higher-dimensional operator, it may constitute sizable corrections in the presence of a high Higgs number density $|\h|^2$, such as in halos. We leave a detailed exploration of these features to future work.

\section{Conclusion}
In this work, we have introduced the Fuzzy Dark Sector paradigm, motivated by theoretical considerations (e.g. the proliferation of fuzzy scalars in the string-axiverse) as well as FDM tensions with observations. As a concrete realization of a FDS, we have extensively studied the fuzzy Abelian-Higgs model. In contrast to a low-energy phenomenological model of multiple ultra-light degrees of freedom, the Abelian-Higgs model offers a UV origin for the signs and strengths of its interactions.

In cosmology, the FDS transfer function is characterized by a single scale, hence mimicking non-interacting single-component FDM. Remarkably, the FDS can mimic a much heavier or much lighter FDM candidate, depending on the model parameters, such as mass hierarchy of the Higgs and gauge field. This apparent degeneracy is lifted by considering the properties of solitons in DM halos. On the other hand, unlike conventional fuzzy dark matter, the FDS admits a striking diversity of solitons, with density profiles strongly depending on the relative mass fraction in the different components of the FDS and on the interaction strengths. Solitons also probe the interactions of the FDS through their stability. The attractive interaction between the Higgs and gauge field leads to a critical soliton mass beyond which solitons are unstable, and which would not be present in the non-interacting theory.

This work represents only the advent of the FDS. There are many possible FDS realizations, each with potentially unique and distinguishing phenomenology. We leave this exploration to future work.\\

\textbf{\textit{Acknowledgements.}}
We thank Katelin Schutz and J. Luna Zagorac for useful comments on the manuscript. C.C. is supported in part by the Arthur B. McDonald Institute via the Canada First Research Excellence Fund and by a Doctoral Research Scholarship
from the Fonds de Recherche du Qu\'ebec--Nature et Technologies. E.M. is supported in part by a Discovery Grant from the Natural Sciences and Engineering Research Council of Canada, and by a New Investigator Operating Grant from Research Manitoba. Kavli IPMU is supported by the World Premier International Research Center Initiative (WPI), MEXT, Japan. EGMF thanks the support of the Serrapilheira Institute.\\

\bibliography{references}

\clearpage
\appendix
\setcounter{equation}{0}
\setcounter{table}{0}
\setcounter{figure}{0}
\renewcommand{\thetable}{A\Roman{table}}
\renewcommand{\thefigure}{A\arabic{figure}}
\renewcommand{\theequation}{A\arabic{equation}}



\end{document}